\documentclass[runningheads]{llncs}
\usepackage[T1]{fontenc}
\usepackage{hyperref}
\usepackage{graphicx,verbatim}
\usepackage{amssymb, amsmath}

\begin{document}
\title{Deep Unrolled Networks in Representation Space Applied to MRI Reconstruction}
\titlerunning{Deep Unrolled Networks in Representation Space}
%

\author{Efe Il\i cak, Baris Imre, Chlo\'e Najac, Ruben van den Broek, Beatrice Lena, Andrew Webb, Marius Staring}  
\authorrunning{Il\i cak et al.}
\institute{Department of Radiology, Leiden University Medical Center, Leiden, Netherlands \\
    \email{e.ilicak@lumc.nl}}
  
\maketitle              
\begin{abstract}
Deep unrolled networks (DUNs) integrate physical forward models with learned regularization in cascaded network architectures, achieving exceptional performance in inverse problems while maintaining interpretability. While most DUNs operate in the object domain (e.g., image space), recent variants explored representation spaces for improved information flow. However, these methods rely on heuristic methods for data consistency (DC), sacrificing fidelity with measurements.

In this work, we introduce DUNE (Deep Unrolled Networks in rEpresentation space), a framework that maintains exact adherence to physical measurements while operating in learned representation spaces. By deriving the DC gradient via the chain rule and implementing it through the Vector-Jacobian Product (VJP), we enable exact backpropagation of measurement residuals into the representation space. This formulation supports diverse architectural backbones, including pre-trained encoders to guide the iterative process. 

We assess DUNE against state-of-the-art baselines on accelerated MRI reconstruction tasks, demonstrating that exact VJP-based gradients yield superior reconstruction quality and structural fidelity across both single-channel portable low-field and multi-channel clinical high-field MRI acquisitions. The code will be available upon publication at \url{https://github.com/EfeIlicak/DUNE}.



\end{abstract}
\section{Introduction}


Inverse problems are central to medical imaging, requiring reconstruction of unknown signals from incomplete or noisy measurements governed by known physical processes~\cite{Fessler}. The ill-posed nature of these problems necessitates regularization to incorporate prior knowledge and ensure stable reconstructions~\cite{Fessler}.

Recently, DUNs have emerged as a powerful paradigm for solving inverse problems, combining the interpretability of model-based iterative algorithms with the representational power of deep learning~\cite{MoDL,Schlemper2019}. By mapping iterative optimization algorithms into cascaded network architectures that embed physical forward models~\cite{E2E}, DUNs provide enhanced interpretability with superior performance. Their effectiveness has been demonstrated across diverse modalities including magnetic resonance imaging (MRI)~\cite{MoDL,Schlemper2019}, low-dose computed tomography~\cite{LDCT}, ultrasound imaging~\cite{DUN-US,US-CT}, and magnetic particle imaging~\cite{DEQ-MPI}. Yet, traditional DUNs operate directly in the object domain (e.g., image or measurement space), where iterative updates compress high-dimensional learned features back into the original signal space at each iteration, creating an information bottleneck that limits representational capacity~\cite{Zhang2022}. 

To alleviate this bottleneck, recent works have explored DUNs in learned representation spaces, including compressed latent spaces~\cite{Lu_2025} and expanded feature spaces~\cite{Zhang2022,Giannakopoulos2024,PGIUN}. However, these approaches rely on heuristic approximations for enforcing data consistency, effectively decoupling the optimization from the underlying physics and sacrificing measurement fidelity.

We introduce DUNE (Deep Unrolled Networks in rEpresentation space), which maintains exact adherence to physical measurements while operating in learned representation spaces. Instead of heuristic encoder projections, we derive the exact gradient via the chain rule and implement it using the Vector-Jacobian Product (VJP). This formulation decouples the encoder from gradient computation, enabling diverse architectural choices including pre-trained models.

Our contributions are: (1) derivation of exact data consistency gradients in representation space via VJP, replacing heuristic approximations; (2) encoder decoupling, enabling the use of pre-trained architectures; (3) validation of our approach on challenging scenarios including accelerated single-channel portable low-field MRI (0.047T) and multi-channel clinical MRI (1.5T-3T) acquisitions.


\section{Background}
\subsection{Problem Formulation}
We consider recovering an unknown image $\mathbf{x} \in \mathbb{C}^N$ from noisy measurements $\mathbf{y} \in \mathbb{C}^M$ governed by:
\begin{equation}
  \mathbf{y} = \mathcal{A}(\mathbf{x}) + \mathbf{n},
\end{equation}
where $\mathcal{A}: \mathbb{C}^N \to \mathbb{C}^M$ is the forward operator and $\mathbf{n}$ is additive noise. This ill-posed inverse problem is typically solved via regularized optimization:
\begin{equation}
  \mathbf{x}^* = \arg \min_{\mathbf{x}} { \frac{1}{2} \| \mathcal{A}(\mathbf{x}) - \mathbf{y} \|_2^2 + \lambda \mathcal{R}(\mathbf{x}) },
\end{equation}
where $\mathcal{R}(\mathbf{x})$ is a regularizer (e.g., total variation) that incorporates priors~\cite{Cukur}.


\subsection{Standard Deep Unrolled Networks}
This optimization problem can be solved iteratively via gradient descent:
\begin{equation}
\mathbf{x}^{(i+1)} = \mathbf{x}^{(i)} - \eta^{(i)} \left[\mathcal{A}^H (\mathcal{A}(\mathbf{x}^{(i)})-\mathbf{y}) + \nabla\mathcal{R}(\mathbf{x}^{(i)})\right],
\end{equation} where $\eta^{(i)}$ is the step size and $\mathcal{A}^H$ is the adjoint operator. 

Alternatively, modern unrolled networks replace the hand-crafted $\mathcal{R}$ with a leaned regularization $\mathcal{R}_\theta$~\cite{MoDL,Schlemper2019}: 
\begin{equation}
\label{image_update}
\mathbf{x}^{(i+1)} = \mathbf{x}^{(i)} - \lambda^{(i)} \mathcal{A}^H (\mathcal{A}\mathbf{x}^{(i)}-y ) - \mathcal{R}_\theta({x}^{(i)}),
\end{equation}
where $\lambda^{(i)}$ is the data consistency weight.


While effective, this formulation compresses learned features back to the object domain (e.g., into two channels for real and imaginary components in MRI reconstruction) at each iteration, creating an information bottleneck~\cite{Zhang2022,Giannakopoulos2024}. Importantly, hybrid-domain variants that operate in both image and measurement domains (e.g., dual image \& k-space networks~\cite{zhou2020dudornet}) suffer from the same limitation, as they still revert to the object domain for computing data consistency updates.


\subsection{Deep Unrolled Networks in Representation Space}
\begin{figure}[tb]
\includegraphics[width=\textwidth]{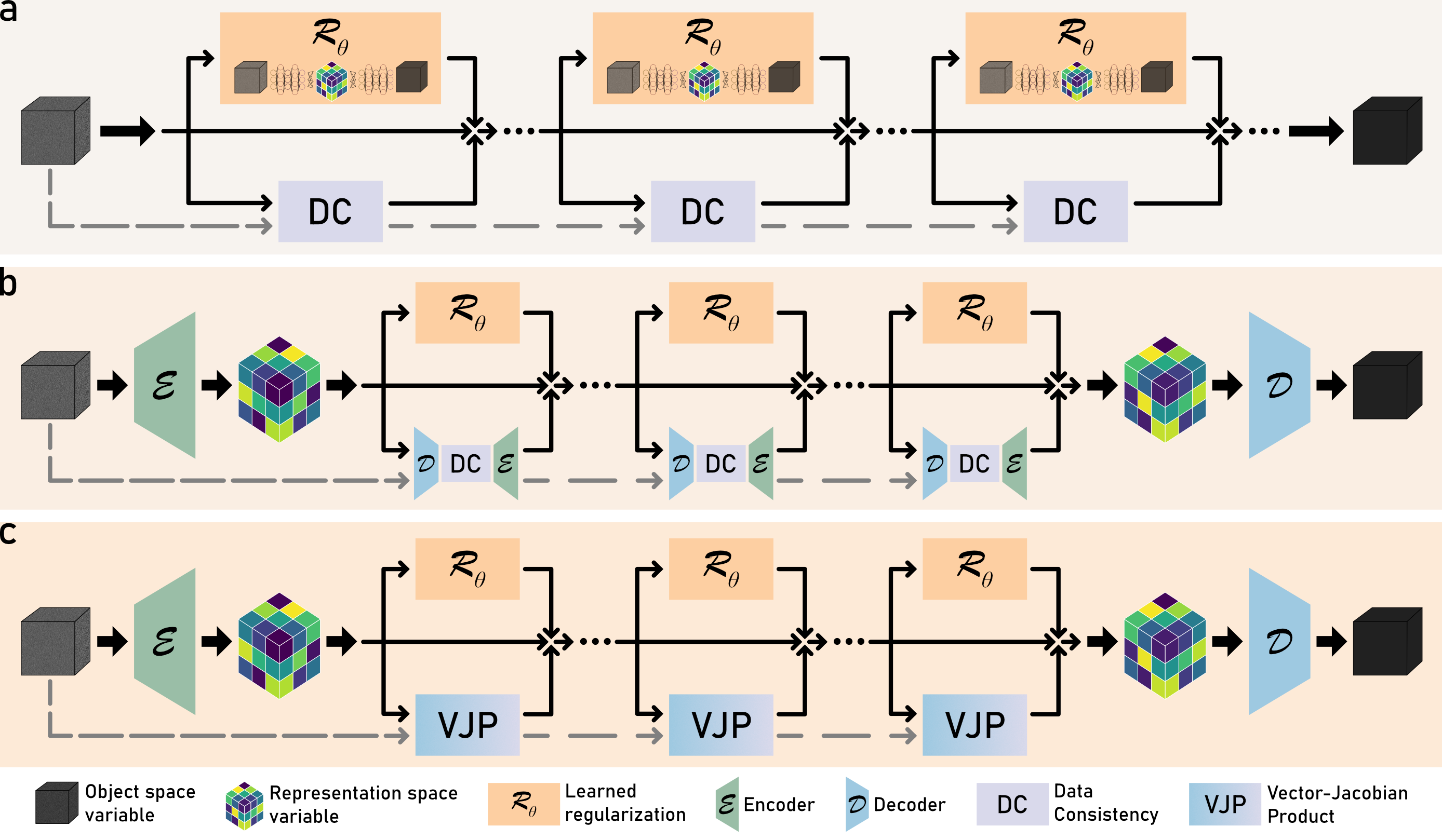}
\caption{Overview of DUN architectures. (a) Standard DUNs use learned regularization $\mathcal{R}_\theta$, but compress learned features back to object space at each iteration. (b) Heuristic representation space DUNs unroll in representation space, but approximate the data consistency (DC) gradients using encoder (${\mathcal{E}}$) projections. (c) DUNE unrolls in representation space with exact  the exact DC gradients computed via Vector-Jacoboian Product (VJP) through decoder (${\mathcal{D}}$), decoupling the ${\mathcal{E}}$ (used only for initialization) from iterative updates.} 
\label{DUNE}
\end{figure}

In this work, we reformulate the inverse problem to perform optimization directly in a learned representation domain. The objective function in this case becomes:
\begin{equation}
    \mathbf{f}^* = \arg \min_{\mathbf{f}} { \underbrace{\frac{1}{2} \| \mathcal{A}(\mathcal{D}(\mathbf{f})) - \mathbf{y} \|_2^2}_{\mathcal{L}_{\text{DC}}(\mathbf{f})} + \lambda \mathcal{R}(\mathbf{f})},
\end{equation}
where $\mathbf{f} \in \mathbb{C}^Q$ is the representation variable, which may be a compressed latent space ($Q<N$) or an expanded feature space ($Q>N$). The mapping is defined by an encoder $\mathcal{E}: \mathbb{C}^N \to \mathbb{C}^Q$ and a decoder $\mathcal{D}: \mathbb{C}^Q \to \mathbb{C}^N$, such that $\mathbf{f} = \mathcal{E}(\mathbf{x})$ and $\mathbf{x} \approx \mathcal{D}(\mathbf{f})$. The $\mathcal{L}_{\text{DC}}(\mathbf{f})$ enforces data consistency (DC) with measurements $\mathbf{y}$, and $\mathcal{R}(\mathbf{f})$ represents regularization in the representation domain.

By applying the chain rule to $\mathcal{L}_{\text{DC}}(\mathbf{f})$, we obtain:
\begin{equation}
    \nabla_\mathbf{f}\mathcal{L}_{\text{DC}} = \left(\frac{\partial \mathcal{D}}{\partial \mathbf{f}}\right)^{H} \nabla_{\mathbf{x}} \left( \frac{1}{2} \| \mathcal{A}(\mathbf{x}) - \mathbf{y} \|_2^2 \right) \bigg|_{\mathbf{x}=\mathcal{D}(\mathbf{f})} = J_\mathcal{D}(\mathbf{f})^{H} \mathbf{g}_x,
\end{equation}
where $\mathbf{g}_x := \mathcal{A}^H (\mathcal{A}(\mathcal{D}(\mathbf{f})) - \mathbf{y})$ is the residual gradient in the object space, and $J_\mathcal{D}(\mathbf{f})$ is the Jacobian of the decoder. Thus, the representation space update is driven by the standard object space DC gradient mapped into the feature space through the adjoint of the decoder’s Jacobian.

\subsubsection{Heuristic Approximation in DUNs} Recent approaches have explored unrolling in both lower-dimensional latent spaces~\cite{Lu_2025} and higher-dimensional feature spaces~\cite{Zhang2022,Giannakopoulos2024,PGIUN} to improve reconstruction quality while maintaining interpretability. However, these methods rely on heuristic gradient approximations for DC, using $\mathcal{E}$ to project residuals back into the representation space:
\begin{equation}
\label{heuristic_update}
    \mathbf{f}^{(i+1)} = \mathbf{f}^{(i)} - \eta^{(i)} \mathcal{E} \left( \mathcal{A}^H (\mathcal{A}( \mathcal{D}(\mathbf{f}^{(i)})) - \mathbf{y}) \right) - \mathcal{R}_\theta(\mathbf{f}^{(i)}).
\end{equation}
However, this approach imposes a dual functional requirement on $\mathcal{E}$. While the exact gradient descent direction in representation space is defined by $J_{\mathcal{D}}(f)^{H} g_x$, the heuristic update (Eq. \ref{heuristic_update}) substitutes this term with $\mathcal{E}(g_x)$. Therefore, $\mathcal{E}$ must simultaneously serve as a feature extractor, and as a proxy for $J_{\mathcal{D}}(f)^{H}$. Some methods~\cite{PGIUN} attempt to improve this approximation by regularizing the encoder-decoder pair toward invertibility, aiming to make $\mathcal{E} \approx \mathcal{D}^{-1}$. Nonetheless, because feature encoding and adjoint mapping are fundamentally distinct operations, the heuristic approach forces a compromise between representational accuracy and physical fidelity.

\subsubsection{Exact Gradient via VJP} To avoid the limitations of heuristic projections, DUNE computes the exact representation space gradient via the Vector–Jacobian Product (VJP). By leveraging standard autograd engines, the term $J_\mathcal{D}(\mathbf{f})^{H} \mathbf{g}_x$ is computed efficiently by backpropagating $\mathbf{g}_x$ through $\mathcal{D}$. This yields the DUNE update rule:
\begin{equation}
\label{exact_update}
    \mathbf{f}^{(i+1)} = \mathbf{f}^{(i)} - \eta^{(i)} \underbrace{J_\mathcal{D}(\mathbf{f}^{(i)})^{H} \mathbf{g}_x}_{\text{VJP}} - \mathcal{R}_\theta(\mathbf{f}^{(i)}).
\end{equation}

\subsubsection{Encoder Decoupling} Note that in Eq. \ref{exact_update}, the DC gradient depends exclusively on $\mathcal{D}$ and $\mathcal{A}$,  while $\mathcal{E}$ serves solely for initialization ($\mathbf{f}^{(0)} = \mathcal{E}(\mathbf{x}^{(0)})$). This structural decoupling removes the need for $\mathcal{E}$ to serve as a gradient projector or an inverse. This allows the integration of high-capacity, pre-trained models to guide the reconstruction process without compromising exact data consistency updates. This is especially valuable in settings where training data is limited.

\section{Experiments}
DUNE was evaluated on a prototype single-channel portable low-field and multi-channel clinical high-field MRI datasets. To showcase structural decoupling, we employed a frozen pre-trained $\mathcal{E}$ for low-field reconstructions, and a trainable $\mathcal{E}$ for the high-field setting. All models were optimized for 100 epochs using Huber loss and the AdamW optimizer on an NVIDIA RTX 4000 Ada GPU. Random undersampling masks were generated per batch at each epoch with random monotonic contrast augmentation for robustness. Reconstruction quality was quantified using PSNR, SSIM, and HaarPSI metrics calculated on foreground regions~\cite{IQA}. Statistical significance was assessed using the Wilcoxon signed-rank test between DUNE and competing methods, with $^*$ denoting $p < 0.01$.

\subsection{Portable Low-Field MRI Reconstruction}
Portable MRI systems operating at low-field strengths ($B_0<0.1$T) using permanent magnet arrays offer accessible point-of-care imaging~\cite{Ayde}. However, compared to clinical systems, they face severe challenges including several orders of magnitude lower signal-to-noise ratios, and limited measurement redundancy from single-channel receiver coils~\cite{Tom}. While DUNs have recently shown promise for portable MRI reconstruction~\cite{shimron2025,DUN-IMG}, the scarcity of large-scale low-field training data remains a critical bottleneck in training task-specific networks. 

\paragraph{DUNE-$f$-pre:} To demonstrate the flexibility of DUNE in such limited-data settings, we propose a feature space model that leverages a frozen pre-trained encoder (DUNE-$f$-pre). To this end, we employ the stem and first stage of ConvNeXt-Tiny model trained on ImageNet-1k for image classification task~\cite{ConvNeXt}, as the encoder $\mathcal{E}$. The decoder $\mathcal{D}$ uses $4\times4$ transposed convolutions to reverse the encoder's spatial downsampling, followed by two pre-activation residual blocks with GELU activation. For the regularization network $\mathcal{R}_\theta$, we utilize three 3D squeeze-and-excitation residual blocks with LeakyReLU activation. The model is unrolled for 3 iterations with a constant channel dimension of 96. 

\subsubsection{Portable MRI Experiments}
We conducted an extensive ablation study comparing DUNE-$f$-pre against: (i) image domain variant; (ii) feature space variant with heuristic gradient updates; (iii) feature space variant with joint reconstruction and autoencoder loss ($\mathcal{L}_{total} = \mathcal{L}_{recon} + \mathcal{L}_{AE}$). These experiments isolate the benefits of representation space reconstruction, exact VJP gradients, and the impact of enforcing $\mathcal{E} \approx \mathcal{D}^{-1}$. 

We also compared DUNE-$f$-pre against state-of-the-art baselines: a 2D iterative shrinkage-thresholding model (ISTA-NET, 10 iterations)~\cite{shimron2025}, and a 3D dual-domain model (DUN-DD, 5 iterations) with parallel Fourier- and image-domain branches that are fused via an attention U-Net~\cite{DUN-DD}.

All models were trained on emulated low-field MRI data~\cite{DUN-IMG} from high-field fastMRI brain volumes~\cite{fastMRI}, using 400 volumes for training, and 80 for validation. Pseudo-low-field measurements were undersampled at $R=2$. The trained model and baselines were then tested on 3D T2-w acquisitions from 9 volunteers obtained with a prototype 0.047T portable MRI system~\cite{Tom} at $R=2$, without any additional training. Volunteer measurements with the prototype system were conducted in accordance with approval from the local ethics committee.

\begin{figure}[tb]
\includegraphics[width=\textwidth]{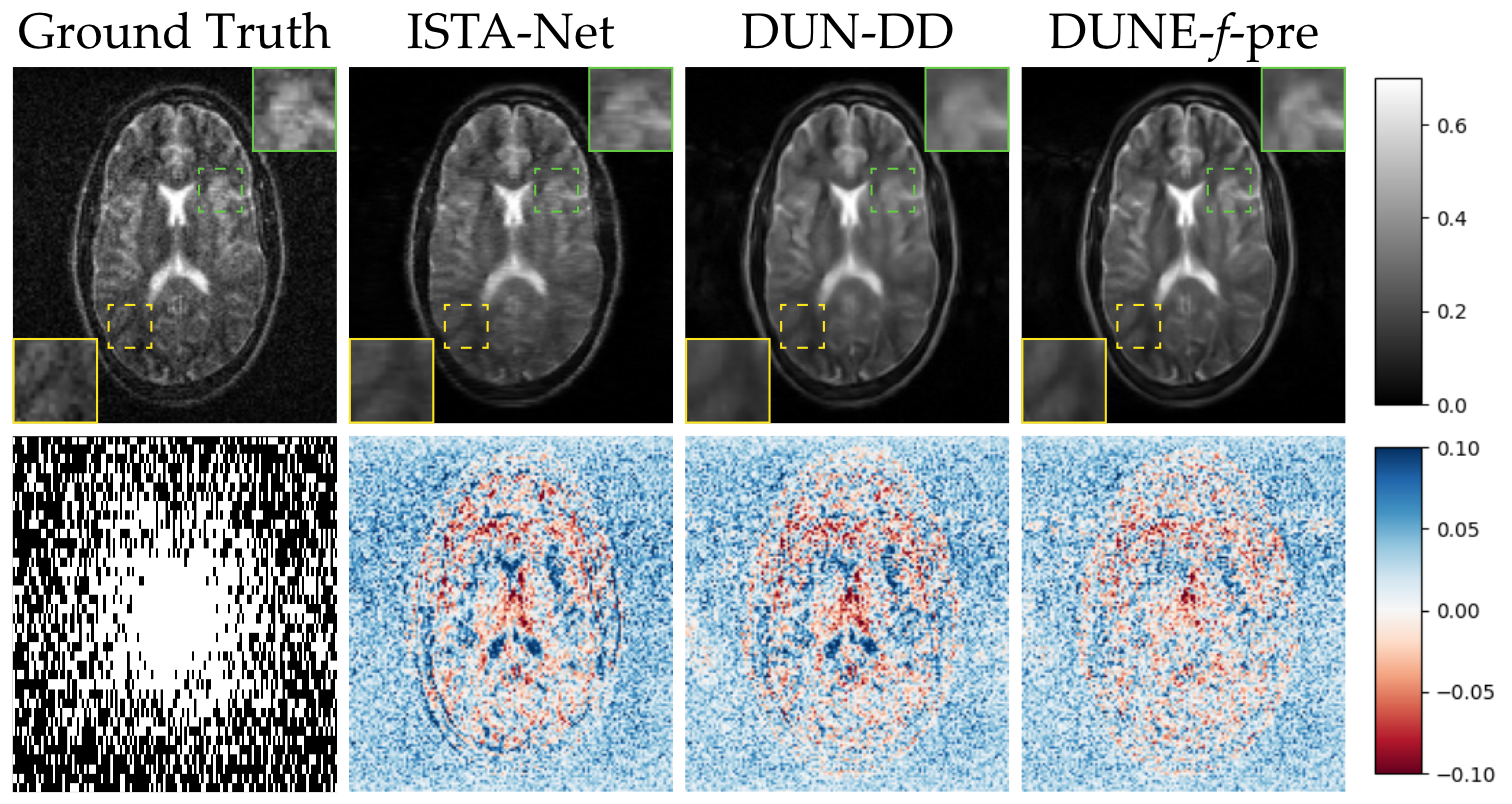}
\caption{Representative T2-w reconstructions obtained with a prototype 0.047T portable MRI scanner. Top: ground truth ($R=1$) and reconstructions at $R=2$ with zoomed regions of interest; bottom: sampling mask and corresponding error maps.}
\label{Fig:VLF}
\end{figure}

\subsection{Clinical High-Field MRI Reconstruction}
We assessed DUNE on clinical high-field MRI reconstruction using 2D fastMRI knee data~\cite{fastMRI}, with 1170 slices for training, 388 for validation, and 776 for testing. Multi-coil acquisitions were undersampled at $R=5$, and all models employed 12 cascades. To evaluate DUNE’s flexibility across different representational domains, we implemented two distinct configurations with learnable $\mathcal{E}$: (i) feature space DUNE (DUNE-$f$), (ii) latent space DUNE (DUNE-$z$). 

\paragraph{DUNE-$f$:} We designed DUNE-$f$ to closely follow Feature-VarNet~\cite{Giannakopoulos2024} for controlled comparison. Following coil sensitivity estimation and combination, images are projected to and from an 18-channel feature space using single $5\times5$ convolutional layers for both $\mathcal{E}$ and $\mathcal{D}$. This model employs a shared encoder-decoder pair with non-shared $\mathcal{R}_\theta$.

\paragraph{DUNE-$z$:} We further propose DUNE-$z$ as a proof-of-concept to demonstrate unrolling in compressed latent spaces. Multi-coil measurements are projected to and from a 12-channel latent representation using single $1\times1$ convolutional layers as $\mathcal{E}$ and $\mathcal{D}$. DC is enforced via the forward model with coil sensitivities, while regularization operates in the latent space. This model employs a shared encoder-decoder pair with a shared $\mathcal{R}_\theta$.
 
\subsubsection{Clinical MRI Experiments} 
We compared DUNE against established object-, feature-, and hybrid feature-image domain models: E2E-VarNet~\cite{E2E}, Feature-VarNet~\cite{Giannakopoulos2024}, and FI-VarNet~\cite{Giannakopoulos2024}. For fair comparison, all models utilize a 4-level UNet with 18 base channels for $\mathcal{R}_\theta$, except FI-VarNet which combines transformers with UNet. Coil sensitivity maps are estimated using a dedicated UNet~\cite{E2E}.

\section{Results}
\subsection{Portable Low-Field MRI Reconstruction}

Table~\ref{Tab:VLF0} presents ablation results on prototype 0.047T portable MRI data, isolating the three key aspects of our approach. DUNE-$f$-pre significantly outperforms both the image domain and the heuristic gradient variant, validating the benefits of feature space reconstruction and exact VJP-based gradients, respectively. Critically, adding an autoencoder loss to enforce $\mathcal{E} \approx \mathcal{D}^{-1}$ degrades performance, confirming that exact gradients eliminate the need for this constraint, which unnecessarily restricts model flexibility.
\begin{table}[tb]
\centering
\caption{Reconstruction performance of DUN variants sharing the same $\mathcal{R}_\theta$ architecture, evaluated at $R=2$ on T2-w measurements from a prototype 0.047T portable MRI. Best results in \textbf{bold}.}\label{Tab:VLF0}
\begin{tabular*}{\textwidth}{@{\extracolsep{\fill}}llccc}
\hline
Ablation Variants & Update Rule & PSNR (dB) & SSIM (\%) & HaarPSI (\%) \\
\hline
Image Domain                            & Eq. \ref{image_update}    & 26.59±1.46$^*$ & 84.55±1.58$^*$ &  66.54±2.77$^*$\\  
Heuristic-$f$-pre                       & Eq. \ref{heuristic_update}&25.45±1.19$^*$ & 83.32±2.32$^*$ &  69.30±2.89$^*$\\ 
DUNE-$f$-pre (+ $\mathcal{L}_{AE}$)  & Eq. \ref{exact_update}    &26.62±1.35$^*$ & 85.28±2.20$^*$ &  70.28±2.39$^*$\\ 
DUNE-$f$-pre                            & Eq. \ref{exact_update}    & \textbf{27.25±1.48{ }}  &  \textbf{86.41±1.89{ }}  &   \textbf{72.36±2.80{ }}  \\ 
\hline
\end{tabular*}
\end{table}

Table~\ref{Tab:VLF} and Figure~\ref{Fig:VLF} compare DUNE-$f$-pre against baseline methods on T2-w images ($R=2$, prototype 0.047T scanner). DUNE-$f$-pre achieves superior reconstruction quality across all metrics with better preservation of anatomical structures, demonstrating effective transfer of pre-trained features to portable low-field MRI.
\begin{table}[tb]
\centering
\caption{Reconstruction performance of competing methods on T2-w acquisitions from the prototype 0.047T portable MRI at $R=2$, including inference time per volume, and trainable model size. Best results in \textbf{bold}.}\label{Tab:VLF}
\begin{tabular*}{\textwidth}{@{\extracolsep{\fill}}lccccc}
\hline
Method & PSNR (dB) & SSIM (\%) & HaarPSI (\%) & Time (s) & Size (M) \\
\hline
ISTA-Net            & 25.26±0.95$^*$ & 79.96±1.86$^*$ &  60.99±3.08$^*$ & 0.60±0.05 & 0.20\\ 
DUN-DD              & 26.04±1.36$^*$ & 83.12±1.80$^*$ &  65.01±3.07$^*$ & 4.43±1.28 & 3.53\\ 
DUNE-$f$-pre        & \textbf{27.25±1.48{ }}  &  \textbf{86.41±1.89{ }}  & \textbf{72.36±2.80{ }}  & 1.95±0.20 & 2.00\\ 
\hline
\end{tabular*}
\end{table}

\subsection{Clinical High-Field MRI Reconstruction}

\begin{figure}[tb]
\includegraphics[width=\textwidth]{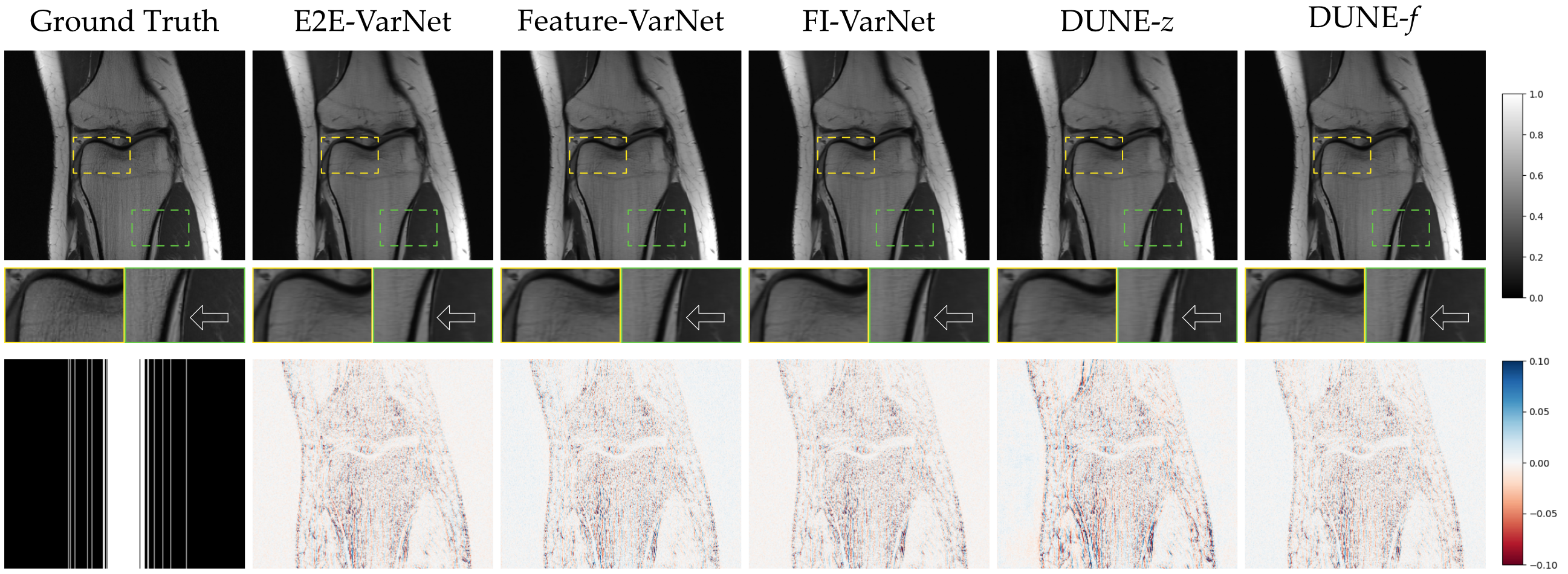}
\caption{Representative reconstructions at $R=5$ from fastMRI knee data. Top: ground truth and reconstructions; middle: zoomed regions of interest; bottom: sampling mask and corresponding error maps.} 
\label{Fig:Knee}
\end{figure}

Table~\ref{Tab:Knee} and Figure~\ref{Fig:Knee} present results on fastMRI knee data at $R=5$. DUNE-$f$ achieves superior performance across all metrics, outperforming object space (E2E-VarNet), feature space (Feature-VarNet), and hybrid (FI-VarNet) baselines. DUNE-$z$ demonstrates competitive performance with substantially reduced model size, illustrating framework flexibility. Inference times are comparable across methods, confirming VJP introduces negligible test-time  overhead.

\begin{table}[tb]
\caption{Quantitative comparisons on the fastMRI knee data at $R=5$, including reconstruction metrics, inference time per slice, and model size. Best results in \textbf{bold}.}\label{Tab:Knee}
\begin{tabular*}{\textwidth}{@{\extracolsep{\fill}}lccccc}
\hline
DUN Model & PSNR (dB) & SSIM (\%) & HaarPSI (\%) & Time (s) & Size (M) \\
\hline
E2E-Varnet~\cite{E2E}                    & 30.27±2.84$^*$ & 72.64±9.63$^*$ & 82.16±4.94$^*$ & 0.09±0.01 & 29.94\\  
Feature-Varnet~\cite{Giannakopoulos2024} & 30.56±2.97$^*$ & 73.44±9.86$^*$ & 83.54±5.41$^*$ & 0.08±0.01 & 30.00\\ 
FI-Varnet~\cite{Giannakopoulos2024}      & 30.56±2.92$^*$ & 73.34±9.77$^*$ & 83.40±5.21$^*$ & 0.13±0.01 & 61.84\\ 
DUNE-$z$               & 29.37±2.69$^*$ & 69.69±9.88$^*$ & 78.79±4.95$^*$ & 0.08±0.01 & 2.94\\ 
DUNE-$f$              & \textbf{30.64±3.04} & \textbf{73.62±9.90} & \textbf{83.90±5.38} & 0.10±0.01 & 29.98\\ 
\hline
\end{tabular*}
\end{table}

\section{Conclusion}
We introduced DUNE, a framework for deep unrolled networks operating in learned representation spaces with exact physical fidelity. By computing DC gradients via VJP, we replace heuristic encoder approximations while decoupling the encoder from iterative updates. This structural property enables flexible encoder design, demonstrated using frozen pre-trained encoders for low-field MRI where training data is scarce, and trainable encoders for high-field MRI, with strong performance in both scenarios.

Evaluations across these challenging scenarios show that DUNE consistently outperforms heuristic methods, while ablation studies confirm that enforcing encoder-decoder invertibility degrades performance. DUNE-$f$ achieves superior reconstruction quality across all baselines, while DUNE-$z$ offers competitive performance with substantially reduced parameters. While demonstrated on MRI, the approach extends naturally to other inverse problems with known forward models.




\bibliographystyle{splncs04}
\bibliography{refs}
\end{document}